\documentclass[reprint,aps,prx,longbibliography,superscriptaddress,floats,floatfix]{revtex4-1}

\usepackage{graphicx}
\usepackage{amsmath}
\usepackage{amssymb}
\usepackage[utf8]{inputenc}

\usepackage{xcolor}
\definecolor{dark-green}{RGB}{0, 128, 0}
\definecolor{dark-red}{rgb}{0.4,0.15,0.15}	
\definecolor{dark-blue}{rgb}{0.15,0.15,0.4}	
\definecolor{medium-blue}{rgb}{0,0,0.5}		

\usepackage{hyperref}
\hypersetup{colorlinks, linkcolor={dark-red}, citecolor={dark-blue}, urlcolor={medium-blue}}

\usepackage{soul}

\begin{document}

\title{Observation of supersymmetric pseudo-Landau levels in strained microwave graphene}
\author{Matthieu Bellec}
\email{matthieu.bellec@inphyni.cnrs.fr}
\affiliation{Universit\'e C\^ote d'Azur, CNRS, Institut de Physique de Nice, France}
\author{Charles Poli}
\affiliation{Department of Physics, Lancaster University, Lancaster LA1 4YB, UK}
\author{Ulrich Kuhl}
\affiliation{Universit\'e C\^ote d'Azur, CNRS, Institut de Physique de Nice, France}
\author{Fabrice Mortessagne}
\email{fabrice.mortessagne@inphyni.cnrs.fr}
\affiliation{Universit\'e C\^ote d'Azur, CNRS, Institut de Physique de Nice, France}
\author{Henning Schomerus}
\email{h.schomerus@lancaster.ac.uk}
\affiliation{Department of Physics, Lancaster University, Lancaster LA1 4YB, UK}

\begin{abstract}
Using an array of coupled microwave resonators arranged in a deformed honeycomb lattice, we experimentally observe the formation of pseudo-Landau levels in the whole crossover from vanishing to large pseudomagnetic field strength. This is achieved by utilizing an adaptable set-up in a geometry that is compatible with the pseudo-Landau levels at all field strengths. The adopted approach enables us to observe fully formed flat-band pseudo-Landau levels spectrally as sharp peaks in the photonic density of states, and image the associated wavefunctions spatially, where we provide clear evidence for a characteristic nodal structure reflecting the previously elusive supersymmetry in the underlying  low-energy theory. In particular, we resolve  the full sublattice polarization of the anomalous 0th pseudo-Landau level, which reveals a deep connection to zigzag edge states in the unstrained case.
\end{abstract}

\maketitle

\section{Introduction}

Topological states enjoy intense attention as they equip quantum systems with desirable robust properties. Much of the early focus rested on their unique spectral positions as isolated or dispersive states in a band gap, as well as their spatial localization at edges and interfaces \cite{Has10}, or more recently also corners \cite{Ben17}. More fundamental characterizations, on the other hand, often invoke a third, somewhat deeper feature of topological states, which is connected to the anomalous expectation values of the underlying symmetry operators  \cite{Nie83,Cal85}.
In momentum space, this feature underpins, e.g., the unidirectional chiral currents around the edges of topological insulators, whilst in real space it manifests itself, e.g., in the sublattice polarization of defect states in bipartite lattice systems \cite{Pol15}, as has been exploited in recent topological lasers or non-linear limiters based on photonic Su-Schrieffer-Heeger structures \cite{StJ17,Zha18,Par18,arXrei19a}.

An additional attractive aspect of these anomalous features are that they tie topological effects together that are often seen as separate, due to the varied nature of the specific encountered  spectral and spatial features that first come into focus. A prime example are flat bands, which have been observed in recent experiments  focussing on Lieb lattices \cite{Guz14,Muk15,Die16,Pol17}
and one-dimensional counterparts \cite{Bab16}, as well as suitably deformed graphene \cite{Lev10} and analogous quantum \cite{Gom12} and classical systems
\cite{Rec13,Wen19}. In the latter case they constitute pseudo-Landau levels arising from a synthetic magnetic field
\cite{Gui10}. In particular, signatures of photonic pseudo-Landau levels have been detected by probing the edges of a honeycomb array of optical waveguides \cite{Rec13}.
A second example are a class of helical edge states in reciprocal systems, as observed, e.g., in zigzag terminated graphene \cite{Fuj96,Sal17}. While these bulk and edge phenomena do not naturally fall into the scope of standard topological band structure theory \cite{Has10}, they are still intimately linked to wavefunctions with a characteristic sublattice polarization. This provides a promising perspective from which one can seek to develop very general unifying descriptions (see, e.g., Ref.~\cite{Kun18} for a recent approach utilizing this perspective).

In this work, we demonstrate experimentally for the case of photonic graphene-like systems that the anomalous edge and bulk phenomena tied to sublattice polarization are in fact directly linked. This is achieved
by tracing the formation of pseudo-Landau levels all the way from vanishing to large pseudomagnetic field strengths.
In particular, we report the direct observation of the spatially resolved sublattice polarization in the 0th pseudo-Landau level of strained photonic graphene, and trace it back to the unstrained case, where the system only possesses edge states. By observing a characteristic nodal structure for the higher-order levels, we can then establish a direct link to the supersymmetric Hamiltonian of the underlying low-energy theory \cite{Jac84}.
Thereby, our observations connect a broad variety of topological phenomena to a unifying principle.

\begin{figure}[t]
	\includegraphics{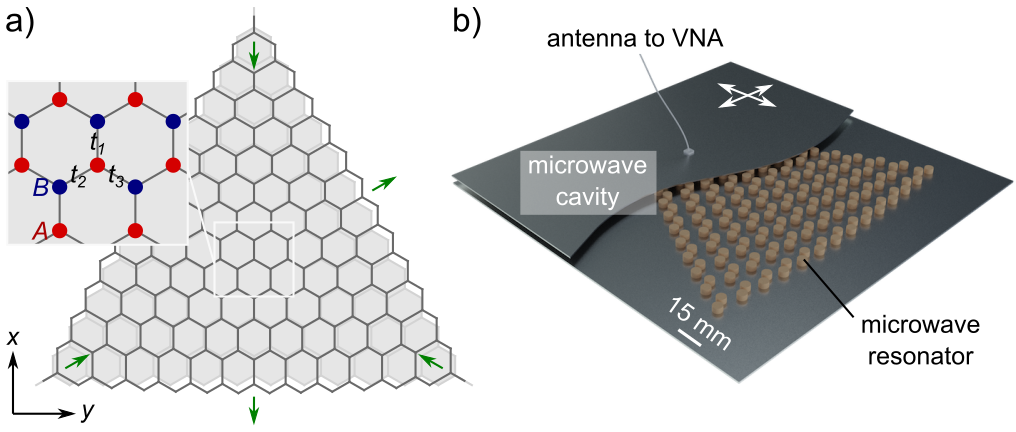}
	\caption{\textbf{Experimental set-up.}
	a) Sketch of the unstrained (gray) and strained (black) honeycomb lattice geometry, forming a zigzag terminated triangle of size $L=14$.
	The green arrows indicates the directions $\boldsymbol{\rho}_l$, $l={1, 2, 3}$,  of the effective triaxial strain.
	Inset: the underlying lattice is composed of two sublattices $A$ (red) and $B$ (blue).
	The local coupling strengths are denoted $t_{l}$.
	They depend on the location of the bonds in the lattice [see Eq.~\eqref{eq:couplingprofile} and the text for details).
	b) In the experiment, the lattice is composed of 196 identically designed cylindrical dielectric resonators that are coupled through the evanescent field of the fundamental TE mode.
	The structure is placed inside a microwave cavity made of two metallic plates (top plate only partially shown).
	A loop antenna, mounted on a scanning system (white arrows) crossing the top plate and connected to a vectorial network analyzer (VNA), is used to generate and collect the microwave signal both spectrally and spatially resolved, which allows to obtain the local density of states in the system.}
	\label{fig:exp}
\end{figure}

\section{Results}

\subsection{Microwave set-up and optimal strain geometry}

Our experimental set-up is illustrated in Fig.~\ref{fig:exp}. The unstrained system forms a honeycomb lattice with nearest-neighbor spacing $a_0=13.9\,\mathrm{mm}$, combining two triangular lattices of A and B sites, where each vertex denotes the position of a dielectric microwave resonator with bare frequency $\omega_0=6.653\,\mathrm{GHz}$, while adjacent resonators are coupled at strength $t_0=21.5\,\mathrm{MHz}$ (for details see the materials and methods section).
This gives rise to a standard graphene-like photonic band structure \cite{Bel13, Bel13a}, with two Dirac cones at the $K$ and $K'$ points in the Brillouin zone. Around these two so-called valleys, at relative momentum $\mathbf{q}$, the low-energy dispersion $\omega(\mathbf{q}) \sim \omega_0\pm v|\mathbf{q}|$ resembles massless relativistic particles moving in two dimensions at velocity $v=3a_0 t_0/2$.

In the deformed system, the indicated couplings $t_l$ depend on the distances to the three neighboring resonators, which we can utilize to create a pseudomagnetic field corresponding to that in strained graphene. Such a field arises when the resonators are displaced non-uniformly, where the positions are selected to give a triaxial spatial coupling profile  \cite{Gui10,Sch13}
\begin{equation}
t_l = t_0[1-(\beta/2a_0^2)\boldsymbol{\rho}_l\cdot \mathbf{r}],
\label{eq:couplingprofile}
\end{equation}
where $\mathbf{r}$ refers to the positions of the links between the coupled resonators in the unstrained system
and the bond vectors $\boldsymbol{\rho}_l$ are pointing along these coupling directions. This coupling profile ensures a constant pseudomagnetic field of strength $\beta$ throughout the whole system.
Theoretically, the system is well described in a coupled-mode theory with nearest-neighbour couplings $t_l$ as given above, so that the eigenfrequencies and mode profiles can be obtained from an effective Hamiltonian $H$.
As any such bipartite system, it then displays a chiral symmetry relative to the central frequency, $\Sigma_z (H-\omega_0) \Sigma_z=-(H-\omega_0)$, where the Pauli-like matrix $\Sigma_z$ acts on the sublattice degree of freedom, hence keeps the amplitudes on A sites fixed but inverts those on the B sites.
The balance of zero modes on the A and B sublattices is given by the signature of this operator \cite{Sut86,Pol17},
\begin{align}
&\#(\mbox{A zero modes})-\#(\mbox{B zero modes})
\nonumber\\
&\quad=\mathrm{tr}\,\Sigma_z=\#(\mbox{A sites})-\#(\mbox{B sites}).
\label{eq:zeromodecount}
\end{align}
These zero modes have frequency $\omega_0$, and
as indicated are localized on a given sublattice. Furthermore, the chiral symmetry dictates that all non-zero modes occur in spectral pairs at symmetric positions $\omega_0\pm \delta\omega$, and have equal intensity on both sublattices.

Importantly, and in contrast to earlier experimental work, we select a triangular geometry and terminate the system with zigzag edges. This ensure a number of beneficial features \cite{Pol14,Rac16}. Particularly relevant for us, the boundary conditions are then compatible with the bulk pseudo-Landau levels at all field strengths; furthermore, a consistent coupling profile can be maintained even at maximal field strength, whose description requires to go beyond the conventional low-energy analogy to magnetic fields with opposite signs in the two valleys. Setting the pseudomagnetic field strength to a fixed value, we then see that the values of the couplings dictated by the profile \eqref{eq:couplingprofile} drop to zero exactly at the terminating edges of a zigzag terminated triangle. The corresponding maximal field strength is $\beta_M=4/L$, where the size parameter $L$ counts the terminating  A sites along each edge.

In the experiments we realize these conditions in a system with 196 resonators, corresponding to a triangle with $L=14$ resonators along each terminating edge (see Fig.~\ref{fig:exp}). Of these, $105$ resonators are on the A sublattice while $91$ are on the B sublattice. This allows us to realize field strength up to $\beta=0.2$, where the extremal couplings still exceed the homogeneous resonator linewidth $\gamma=1.7\,\mathrm{MHz}$, and sufficiently close to $\beta_M$ to clearly demonstrate the detailed features of well-formed pseudo-Landau levels.
To obtain the analogous orbital effects for an electron in graphene, a magnetic field of $42000\,\mathrm{T}$ would have to be applied.

\begin{figure}[t]
	\includegraphics{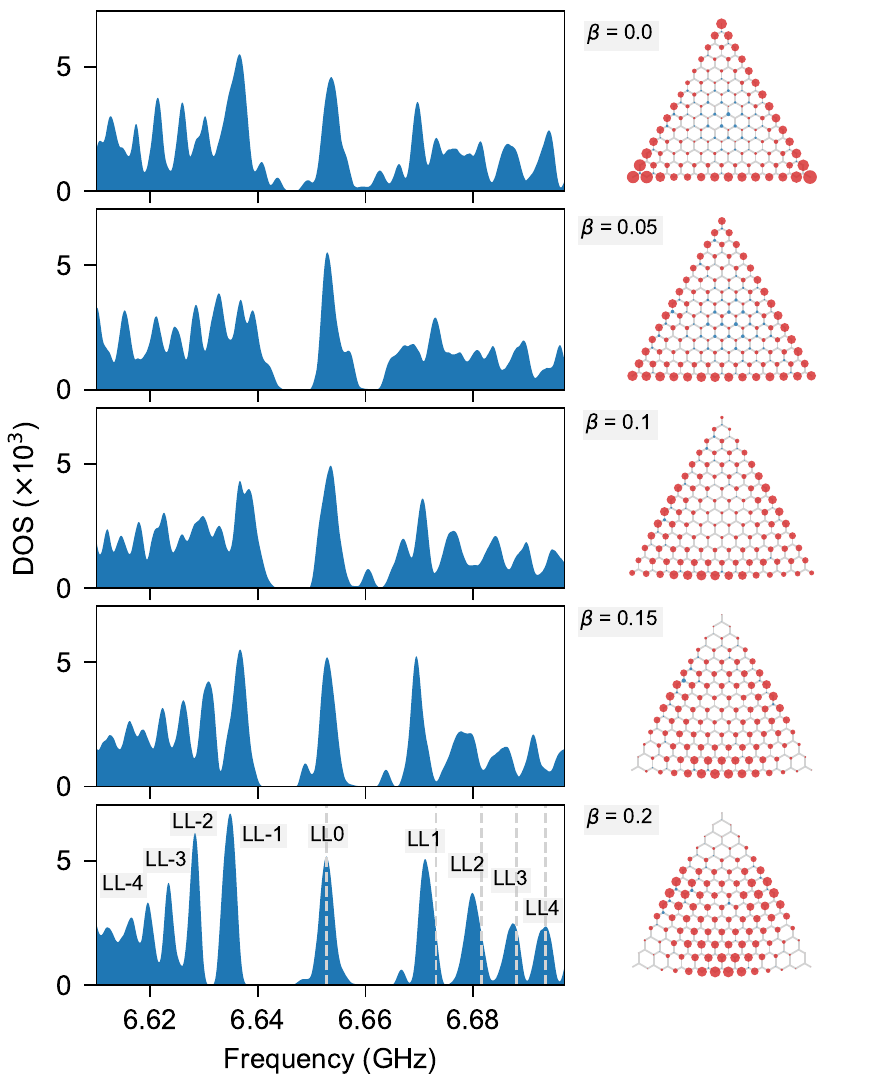}
	\caption{\textbf{Landau-level formation.} Experimentally determined density of states (left panels) and spatially resolved mode intensity associated with the central peak centered at $\omega_0=6.653\,\mathrm{GHz}$ (right panels, obtained by integrating the local density of states over the peak).
	The area of the circles corresponds to the intensity on the A sites (red) and B sites (blue).
	From top to bottom, the pseudomagnetic field strength $\beta$ varies from 0 to 0.2.
	The gray dashed lines in the total density of states for $\beta=0.2$ depict the expected pseudo-Landau levels frequencies from coupled-mode theory.
	The formation of the 0th pseudo-Landau level proceeds continuously by transforming zigzag edge states into bulk states, whilst retaining the degeneracy (spectral weight) and sublattice polarization (remaining confined on the A sublattice).
}
	\label{fig:spectra-LL0}
\end{figure}

\subsection{Formation of pseudo-Landau levels}
Figure \ref{fig:spectra-LL0} shows our main experimental results. The panels on the left show the density of states for the resonator geometry on the right, where each row corresponds to a different value of the pseudomagnetic field strength $\beta$. The color density plot overlayed with the resonator lattice depicts the local density of states integrated over the central peak, situated at the bare resonator frequency $\omega_0$.

In the pristine system ($\beta=0$), this peak arises from the zigzag edge states, which are localized on the terminating resonators. Note that these resonators all occupy the same sublattice of A sites. On this sublattice, the zigzag states decay into the bulk of the system, whilst they maintain a vanishing density on the B sublattice. The simple rule \eqref{eq:zeromodecount} can be exploited to count the number of these zigzag states: As a consequence of the chiral symmetry of the system, this number is expected as the difference $105-91=14=L$ of A and B sites in the system.
Away from the central peak, the density of states displays a broad continuum, with fluctuations arising from the finite-size quantization of bulk graphene-like states, including the states near the Dirac cones.

As the pseudomagnetic field strength is increased, we observe that the spectral weight from this continuum reorganises into a sequence of well-defined peaks, of similar width and weight as the central peak, whose weight and position remains essentially unchanged. The spatial profile of the associated zero-modes, however, changes significantly, in that they move into the bulk, where they form the desired 0th pseudo-Landau level.

\begin{figure*}[t]
	\includegraphics[width=0.9\textwidth]{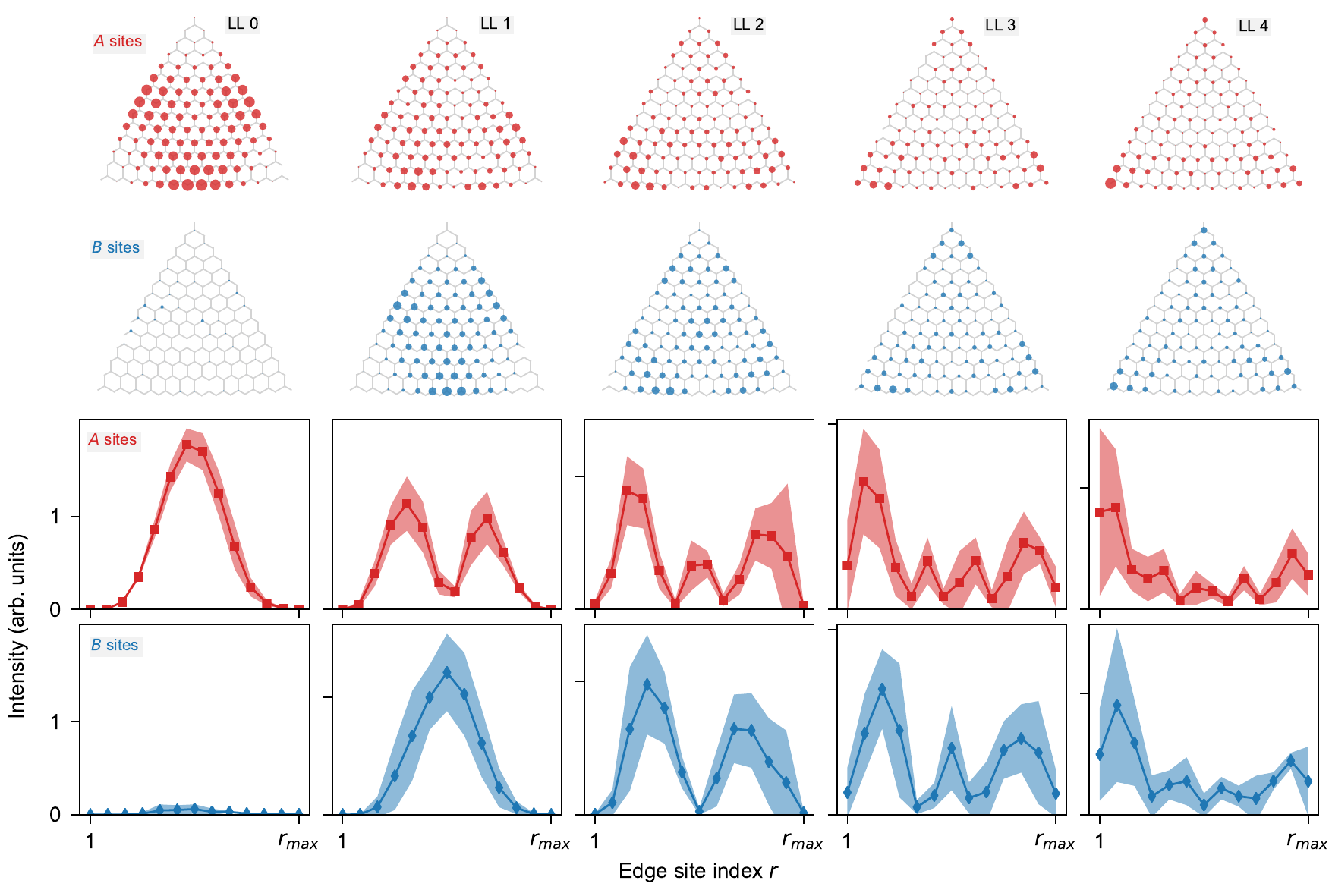}
	\caption{\textbf{Supersymmetric node structure.}
	Experimentally determined spatially resolved mode intensities associated with the pseudo-Landau levels of index $n=0$ to 4 (from left to right) for a large pseudomagnetic field strength $\beta=0.2$.
	Top rows: the area of the  circles correspond to the intensity on the A  sites (above) and B sites (below). The 0th level is almost entirely localized on the A sublattice, while all other levels have almost-equal overall weight on both sublattices.
	Bottom rows: intensity on  A sites (above) and B sites (below), averaged along the three extreme edges at resonator position $r$. Note that $r_{\rm max} = 14$ (resp. 13) for A (resp. B) sites.
	For the higher-order levels, the average is performed combining the indices $\pm n$.
	The shaded area corresponds to the standard deviation.
	The two $y$-axis ticks indicated intensity values of 0 and 1 in arbitrary, but uniformly applied units.
	The pseudo-Landau levels display a clear nodal structure offset by 1 mode index, as further discussed in the text.}
	\label{fig:beta-node}
\end{figure*}

The observed spectral positions of the emerging higher-order pseudo-Landau levels conform well with the characteristic square-root dependence on relativistic Landau levels  \cite{Jac84,Cas09}, revisited later in the text and depicted by the gray dashed lines in Fig.~\ref{fig:spectra-LL0}. The same applies to the observed spectral weights.
As the transformation from the zigzag edge states to the 0th pseudo-Landau level is continuous, this bulk level retains the same spectral degeneracy, hence here consists of $L=14$ modes. The coupled-mode theory predicts that the $n$th pseudo-Landau level encompasses $14-|n|$ states, which explains the gradual drop of the observed spectral weights of the peaks moving outwards from the central peak. Note that this implies an important difference to Landau levels arising from a magnetic field, for which the degeneracy is dictated by the sample area $\propto L^2$, but not the linear size $L$, as observed here and underpinned by general theory \cite{Pol14,Rac16}.

\subsection{Supersymmetric nodal profiles}

As anticipated, a notable feature in the  formation of the 0th pseudo-Landau level verified in the experiment is the observation that across the whole transition, the associated modes remain localized on the A sublattice.
In contrast, the modes in the emerging higher-order levels, whilst also localized in the bulk, are anticipated to display an equal weight on both sublattices. We analyze this distinction in detail in Fig.~\ref{fig:beta-node}.
The top rows display the experimental local density of states in the spectral range of the pseudo-Landau levels for the indices $n=0,1,2,3,4$, all taken at the large pseudomagnetic field strength $\beta=0.2$. As above, the level with $n=0$, shown in the left panels, is located on the A sublattice. The higher-order levels shown in the other panels indeed display an approximately equal weight on both sublattices. Furthermore, they have an intensity profile that increasingly seeps into the corner areas of the triangle, which remains in line with the predictions of  coupled-mode theory where the states form a complete basis.

The other two rows in Fig.~\ref{fig:beta-node} show the local density of states along the edges of the system separately for the A and B sublattice, where we averaged over the three edges and indicate the range of observed values by the shaded areas. For the higher-order levels, we furthermore include the levels with index $-n$ into the average.
Along each edge, we observe standing-wave patterns
\begin{equation}
\psi_r^{(\mathrm{A,\,edge})}\propto \sin [ (|n|+1)\pi r/(L+1)]
\label{eq:wavea}
\end{equation}
for resonator index $r=1,2,3,\ldots, L$ on the A sublattice, and a corresponding pattern
\begin{equation}
\psi_r^{(\mathrm{B,\,edge})}\propto \sin ( |n|\pi r/L)
\label{eq:waveb}
\end{equation}
for resonator index $r=1,2,3,\ldots, L-1$ on the B sublattice.

Note that the mode index $(|n|+1)$ vs $|n|$ in these patterns for the two sublattices is offset by one.
This observation finds its counterpart in the low-energy theory of the pseudo-Landau levels relative to the central frequency $\omega_0$ \cite{Jac84,Cas09,Sch13}, which are described by an effective Hamiltonian
\begin{equation}
H_{0}=\begin{pmatrix} 0 & \pi^\dagger \\
\pi & 0 \end{pmatrix},
\end{equation}
where the Landau-level creation operators fulfill $[\pi,\pi^\dagger]=2v^2\beta/a_0^2$ (we assume $\beta>0$; for opposite deformation the operators $\pi$ and $\pi^\dagger$ interchange their roles of creation and annihilation operators). This corresponds to the Hamiltonian of a relativistic electron in a magnetic field, which is formally identical to a supersymmetric harmonic oscillator  \cite{Jac84,Tha13}. The link to the standing-wave patterns \eqref{eq:wavea} and \eqref{eq:waveb} along the edges arises from the fact that $H_{0}^2=\mathrm{diag}(\pi^\dagger \pi,\pi\pi^\dagger)$ factorizes into two equidistant level sequences $E_n^2=2v^2n \beta/a_0^2$ on the two sublattices, where $n=0,1,2,3\ldots$ on the A sublattice while on the B sublattice $n=1,2,3\ldots$ so that the sequence is offset by one.
While this low-energy analogy does not capture the difference in degeneracies and the importance of boundary conditions described above, it is notable that the offset between the level sequences persists in the experimental observations, and is furthermore reflected in the mentioned nodal wave-patterns.

\section{Discussion}

We achieved the direct observation of the formation of pseudo-Landau levels in deformed honeycomb systems, both spectrally as well as in terms of their key spatial features.
In particular, adopting a flexible dielectric-resonator array design with a purposefully selected geometry allowed us to follow the formation of these levels in the transition from vanishing to large pseudomagnetic field strengths. In this way, we could observe how the 0th pseudo-Landau level originates from the transformation of zigzag edge states into bulk states, whilst maintaining its characteristic anomalous polarization on only one of the two sublattices in the system.
Extending these considerations to the higher-order levels allowed us to
reveal a characteristic nodal structure of the pseudo-Landau level sequence that reflects the supersymmetric structure of the underlying low-energy description. These features underline the general usefulness to account for anomalous expectation values to provide a more general perspective on topological states.

Resolving the reported features in an experiment poses a significant challenge. In electronic systems such as graphene, spectral imaging techniques do not provide the required atomistic resolution, so that the information can only be extracted indirectly, for example from the Fourier transformation of form factors \cite{Dai95,Bos06}. A better resolution is offered by photonic systems \cite{Rec13}, which however so far could not access the characteristic spatial features of the pseudo-Landau levels in the bulk of the system, and furthermore considered a geometry that does not enable to observed fully formed flat-band pseudo-Landau levels \cite{Pol14}.
Drawing on an acoustic analogue \cite{Abb17}, a recent experiment \cite{Wen19} managed to excite a compacton-like state in the 0th Landau level, demonstrating its characteristic sublattice polarization in the bulk \cite{Sch13}.
Here, we exploited an adaptable dielectric microwave-resonator array geometry to provide a complete characterization of the system from the unstrained to the fully strained case.
This allows us to reveal how zigzag edge states transform into the bulk states of the anomalous 0th pseudo-Landau level, where they retain their characteristic sublattice polarization.
Furthermore, this perspective dictates a natural geometry in which maximal pseudomagnetic fields can be attained, and for which the pseudo-Landau levels remain compatible with the boundary conditions at all field strengths, in contrast to the  previous experiments. This is required to obtain pseudo-Landau levels that are flat, which we evidence spectrally by the observation of sharp peaks in the photonic density of states, and enables to reveal the supersymmetric signatures in the nodal structure.

Our results extend to a wide variety of flat-band systems, such as the rich physics arising from higher-order resonator modes in deformed honeycomb lattices, as reported for exciton polaritons in Ref.~\cite{Amo20}.
Practically, the observed spatial features should help to pave the way to applications such as flat-band lasers \cite{Sch13,Pol17,Smi19}, as well as sublattice-dependent sensors where one could exploit that chiral-symmetry breaking perturbations equip the 0th Landau level with a finite weight on the opposite sublattice.

\section{Materials and methods}

\subsection{Experiment}

The experimental setup is designed to realize a microwave system that is well approximated by a nearest-neighbour tight-binding description \cite{Bel13a}. The sites of the lattice are occupied by dielectric microwave resonators with a cylindrical shape and made of ZrSnTiO ceramics (Temex-Ceramics, E2000 series: 5\,mm height, 8\,mm diameter and a refractive index $n\approx 6$) sandwiched between two metallic plates at a distance $h=16\,\mathrm{mm}$. Each resonator supports a fundamental TE mode of bare frequency $\omega_0$ = 6.653\,GHz, which corresponds to the on-site energy of atoms in a tight-binding model. As the resonance frequency is below the cut-off frequency of the first TE mode defined by the two plates the adjacent resonators are coupled through  evanescent wave components leading to an approximately exponential decay of the coupling strength $t$ with the distance between the resonators \cite{Bel13a}. The system is excited via a loop antenna fixed in the movable top plate thus allowing to scan spatially the magnetic field $B_z$, which is the only magnetic field component for this mode \cite{arXrei19b}. From the reflection measurements performed by a vector network analyzer (ZVA 24 from Rohde \& Schwarz) the local density of states can be extracted (for details see Ref.~\cite{Bel13a}) and finally, by integrating over space, the density of states. In all these experiments, we face an intrinsic on-site disorder of $\sim 0.15\%$ in the values of $\omega_0$.

\subsection{Modelling}
For the design of the system, we modelled finite strained photonic honeycomb lattices with a range of system sizes and boundary geometries within coupled-mode tight-binding theory, in which we accounted for the fundamental TE mode and incorporated the experimental distance dependence of the coupling strengths following Ref.~\cite{Bel13a}. Exact diagonalization gives us access to the resonant modes and their spatial intensity distribution.
The modelling confirmed that the triaxial coupling profile \eqref{eq:couplingprofile} can be attained by suitable positioning of the resonators, resulting in an excellent match with the continuum theory predictions for the lowest Landau levels in systems as used in the experiments. The modelling further confirmed that only for zigzag boundaries one can attain the full sublattice polarization of the zeroth Landau level, and only for a triangular shape the Landau levels become maximally degenerate.

\section{Acknowledgements}

MB, UK and FM are grateful to G. Salerno and I. Carusotto for stimulating discussions during their visit to Nice. The authors acknowledge funding by EPSRC via Grants No. EP/J019585/1, EP/P010180/1, and
Programme Grant No. EP/N031776/1, and CNRS through a visiting fellowship of HS to Nice.

\section{Author contributions}
MB and FM carried out the experiments and analyzed the data with input from UK. CP modelled the system with input from HS, who provided the theory and initiated the project. All authors contributed substantially to this work, including the interpretation of the results and the preparation of the manuscript.

%

\end{document}